\documentclass[preprint,aps,12pt,preprintnumbers,eqsecnum,nofootinbib,superscriptaddress]{revtex4}
\usepackage{tabularx} 
\usepackage{amsmath}  
\usepackage{graphicx} 
\usepackage[margin=1in,letterpaper]{geometry} 
\usepackage{physics}
\usepackage{dsfont}
\usepackage{amsfonts,mathrsfs}

\usepackage[final]{hyperref} 
\hypersetup{
	colorlinks=true,       
	linkcolor=blue,        
	citecolor=blue,        
	filecolor=magenta,     
	urlcolor=blue         
}

\newcommand{\qphi}{{\Phi}}
\newcommand{\qtheta}{{\Theta}}
\newcommand{\qpsi}{\Psi}

\newcommand{\bx}{{\mathbf x}}
\renewcommand{\u}{u}
\newcommand{\nn}{n}
\newcommand{\cL}{{\mathcal L}}
\newcommand{\cT}{{\mathscr T}}
\newcommand{\cTT}{{\mathcal T}}
\newcommand{\ZZ}{{\mathbb Z}}
\renewcommand{\chi}{{\mathbf \varphi}}
\newcommand{\Chi}{{\mathbf \Phi}}

\newcommand{\hPhi}{{\hat\Phi}}
\newcommand{\hPsi}{{\hat\Psi}}

\newcommand{\ra}{\rightarrow}
\newcommand{\ba}{{\mathbf a}}
\newcommand{\bA}{{\mathbf A}}
\newcommand{\be}{{\mathbf e}}

\begin{document}

\title{The Universal  Time-Dependent Ginzburg-Landau theory}
\author{Anton Kapustin}
\affiliation{California Institute of Technology}
\author{Luke Mrini}
\affiliation{The College of William \& Mary}

\begin{abstract}
We study the hydrodynamics of superconductors within the framework of Schwinger-Keldysh Effective Field Theory. We show that in the vicinity of the superconducting phase transition the most general  leading-order EFT satisfying the local Kubo-Martin-Schwinger condition is described by a version of the  Time-Dependent Ginzburg-Landau (TDGL) equations augmented with stochastic terms. This version of TDGL is applicable in the gapless regime independently of any microscopic details. Within this approach, it is possible to include systematically the effects of non-uniform temperature and heat conductivity, as well as explicit or spontaneous breaking of time reversal. We also introduce a thermal version of the Josephson relation and use it to construct an exotic hydrodynamics describing a phase of matter where heat can flow without dissipation.
    
\end{abstract}

\maketitle


\section{Introduction} \label{intro}

The celebrated Ginzburg-Landau theory \cite{GinzburgLandau} is extremely successful at explaining equilibrium macroscopic and mesoscopic properties of superconductors. Near $T=T_c$, this success is explained by the observation that the Ginzburg-Landau theory is a universal Effective Field Theory (EFT) of a phase transition to a phase which spontaneously breaks $U(1)$ symmetry.
In this regime it can also be derived from the microscopic BCS theory \cite{Gorkov}. Away from $T_c$, Ginzburg-Landau theory is a useful phenomenological model.  

To describe transport and other non-equilibrium properties of superconductors, time-dependent versions of the Ginzburg-Landau theory (TDGL) are widely used. There exists a large body of work devoted to the derivation of TDGL from the microscopic theory. While early derivations \cite{AbrahamsTsuneto,Schmid} have been criticized by Gor'kov and Eliashberg \cite{GorkovEliashberg}, the latter authors showed that for gapless dirty superconductors  a variant of TDGL  follows from BCS theory. Later more complicated versions of TDGL theory have been deduced from BCS theory under weaker or different assumptions, see e.g. \cite{EbisawaFukuyama,AmbegaokarSchon,Hu,Kopnin,GulianZharkov}. None of the proposed derivations  appear to include the effects of finite heat conductivity.\footnote{While there are many studies of thermoelectric effects in superconductors which make use of TDGL, see e.g. \cite{UllahDorsey1,UllahDorsey2,Ussishkinetal,mukerjeehuse,LVreview}, they typically avoid dealing with inhomogeneous temperature by appealing to Onsager reciprocity. For example, instead of the Nernst effect, one computes its  reciprocal (the Ettingshausen effect), etc. In Ref. \cite{vishfisher} the effects of superconducting fluctuations on heat conductivity is studied using the  phenomenological "Model C" of Ref. \cite{hohenberghalperin}.} It is also unclear if TDGL applies to superconductors whose microscopic description is not given by the standard BCS theory. 
In addition, the ``improved'' versions of TDGL are not controlled expansions in a small parameter in the spirit of effective field theory. For all these reasons, the original TDGL equations of Refs. \cite{AbrahamsTsuneto,Schmid,GorkovEliashberg} remain a popular tool, but are often viewed as a phenomenological model rather than a controlled approximation with a well-defined region of validity.  
 
In this paper we re-examine the theoretical status of TDGL theory from the point of view of effective field theory. Our goal is to derive the most general description of superconductors which are in local thermodynamic equilibrium. In other words, we study the hydrodynamics of superconductors. 
The hydrodynamic description applies whenever the coherence length is larger than the inelastic scattering length (this regime is usually referred to as ``gapless superconductivity''). In particular, it applies right above $T=T_c$ where TDGL is traditionally used to study fluctuation effects in superconductors. 

Recently, a much better understanding of general principles constraining hydrodynamic equations has been achieved \cite{Kovtunetal,HarderKovtunRitz,Haehletal,Haehletal2,Jensenetal,Siebereretal,CrossleyGloriosoLiu,CrossleyGloriosoLiu2}. These developments build on an old observation that the stochastic dynamics of classical dissipative systems can be described using the path-integral formalism with a doubled number of variables. Such a theory can be thought of as a version of the Schwinger-Keldysh (SK) formalism which is used to describe quantum dynamics of arbitrary mixed states. To obtain an effective long-distance low-frequency description, one needs to impose all the relevant symmetries and to Taylor-expand the SK action in the number of spatial and time derivatives. The key new observation is that the requirement of local thermal equilibrium is equivalent to a certain $\ZZ_2$ symmetry of the SK action called the Kubo-Martin-Schwinger (KMS) symmetry \cite{Siebereretal,CrossleyGloriosoLiu,CrossleyGloriosoLiu2} (see \cite{GloriosoLiu} for a pedagogical review). Imposing KMS symmetry ensures that the theory satisfies the usual physical requirements: the existence of the local entropy function and entropy current, Onsager reciprocity, and the fluctuation-dissipation theorem. Positivity of the entropy production is equivalent to the positivity of the imaginary part of the SK action. 

We derive a stochastic hydrodynamic theory of superconductors relying only on such general principles. In fact, we describe two different stochastic
superconducting EFTs. EFT-I is a version of TDGL theory 
and applies near $T=T_c$. At leading order in the derivative
expansion and after dropping stochastic terms and particle-hole symmetry-breaking terms, it agrees
with the Gor'kov-Eliashberg equations. This is rather
remarkable, since the microscopic derivation of the
Gor'kov-Eliashberg equations requires a number of assumptions, including the applicability of the ``dirty
limit''.\footnote{Although it has been proposed in \cite{KW} that the assumption of local equilibrium should suffice.} We also show that at leading order in the derivative expansion there is a unique non-dissipative coupling which violates particle-hole symmetry.

EFT-II is appropriate in the London limit, where Abrikosov vortices are either absent or pinned. Neglecting fluctuations, this EFT is a slight generalization of the transport equations written down by Luttinger \cite{Luttinger}. The main advantage of the Schwinger-Keldysh approach is that the First and Second Laws of Thermodynamics and the fluctuation-dissipation theorem are built in.

In our approach it is straightforward to continue the expansion to higher orders in derivatives (and in the case of EFT-I, to higher orders in $T-T_c$). It is also possible, although less straightforward, to include the effects of a non-uniform temperature. Within the Schwinger-Keldysh formalism, one needs to introduce a dynamical field $\tau$ whose time derivative is the local temperature, $T=\partial_0\tau$. This is a thermal analog of the Josephson relation $\mu+a_0=\partial_0\phi$ between the local electrochemical potential $\mu+a_0$ and the condensate phase $\phi$. The thermal Josephson relation has other interesting uses. As discussed below, it plays a key role in ``superthermal'' hydrodynamic theories describing exotic phases of matter where heat can flow without dissipation. It also allows one to sharpen the well-known analogy between classical mechanics and thermodynamics.

The content of the paper is as follows.
In Section \ref{sec2} we review the SK formalism and illustrate it by deriving a simple  two-fluid model for a superconductor. In Section \ref{EFTI} we derive EFT-I which applies near $T=T_c$ (but without including the effects of non-uniform temperature). In Section \ref{EFTII} we derive EFT-II which applies in the London limit. This EFT includes heat conductivity and thermoelectric effects. We also show how to include such effects in EFT-I. In Section \ref{timereversal} we outline how to incorporate the effects of spontaneous or explicit breaking of time-reversal into the Schwinger-Keldysh formalism. We compare our results with the existing literature in Section \ref{discussion}. In particular, we argue that in clean superconductors the effects of particle-hole symmetry breaking can be much larger than previously believed. In Appendix \ref{appA} we describe some details of the derivation of EFT-I, including a brief discussion of next-to-leading order terms in the expansion. In Appendix \ref{appB} we show how the thermal Josephson relation clarifies the analogy between thermodynamics and mechanics. In Appendix \ref{appC} we construct a hydrodynamic description of an exotic ``superthermal'' phase.

\section{Schwinger-Keldysh EFT and the KMS symmetry} \label{sec2}

Within the Schwinger-Keldysh formalism (see \cite{GloriosoLiu} for a review), each physical degree of freedom $\chi(t)$ gives rise to two variables in the path-integral, the forward-propagating $\chi_1(t)$ and the backward-propagating $\chi_2(t)$. The global symmetry group $G$ is doubled too, so that the total symmetry group is $G_1\times G_2$, where $G_i$, $i=1,2,$ is isomorphic to $G$ and acts only on $\chi_i$. In the classical limit $\chi_1(t)\simeq\chi_2(t)$, and it is more convenient to work with the average $\chi=\frac12(\chi_1+\chi_2)$ and the ``noise'' variable $\Chi=\chi_1-\chi_2$. Accordingly, the symmetry contains the diagonal subgroup $G_D$ which consists of elements of the form $(g,g)\in G_1\times G_2$, $g\in G$. If $G$ is abelian (as will be the case in this paper), then there is also an anti-diagonal subgroup $G_A$ consisting of symmetries of the form $(g,g^{-1})$. 

The action $I[\chi,\Chi]$ for a Schwinger-Keldysh EFT must satisfy the following requirements \cite{GloriosoLiu}:
\begin{enumerate}
    \item $I^*[\chi,\Chi]=-I[\chi,-\Chi]$
    \item $I[\chi,\Chi=0]=0$
    \item ${\rm Im}\, I[\chi,\Chi]\geq 0$
    \item $I[\chi,\Chi]$ is invariant under $G_1\times G_2$
\end{enumerate}
In the classical limit where $\Chi$ is small it is sufficient to expand $I[\chi,\Chi]$ to quadratic order in $\Chi$. Then conditions 1 and 2 say that the part linear in $\Chi$ is real while the quadratic part is imaginary. Condition 3 in addition requires the quadratic part to be $i$ times a positive expression.

If the system is in a local thermal equilibrium, the real and imaginary parts of $I$ are related by the fluctuation-dissipation theorem. If a time-reversal symmetry is present, this condition can be formulated as invariance of $I$ under a $\ZZ_2$ symmetry known as KMS symmetry \cite{Siebereretal,CrossleyGloriosoLiu,CrossleyGloriosoLiu2}. Let $\cT$ be the time-reversal transformation. Then the KMS symmetry acts as follows:
\begin{align}\label{KMSgeneral}
&R_{KMS}(\chi)=\cT(\chi),\\
&R_{KMS}(\Chi)=\cT(\Chi+iT_0^{-1}\partial_0\chi).
\end{align}
Here $T_0$ is the temperature which for now is assumed to be constant. 

Since the KMS symmetry is a $\ZZ_2$ symmetry, constructing KMS-invariant actions seems  straightforward: for any $X(\chi,\Chi)$, the expression $Y(\chi,\Chi)=\frac12(X+R_{KMS}(X))$ is KMS-invariant. However, $Y$ does not necessarily satisfy condition 2. One can try to fix this by replacing $Y(\chi,\Chi)$ with $Z(\chi,\Chi)=Y(\chi,\Chi)-Y(\chi,0)$, but this might destroy KMS-invariance. 

To analyze this issue, we will assume that $\chi$ and $\Chi$ transform linearly and in the same way under $\cT$ (this is true in all examples we are aware of). We have $Y(\chi,0)=\frac12 X\left(\cT(\chi),\cT(iT_0^{-1}\partial_0\chi)\right)$. Note also that $\partial_0\chi$ transforms under $\cT$ with an additional minus sign compared to $\Chi$. Thus if $X$ is linear in $\Chi$ and $\cT$-even, then $Y(\chi,0)$ is $\cT$-odd and thus is not KMS-invariant. On the other hand, if $X$ is $\cT$-odd, then $Y(\chi,0)$ is $\cT$-even and KMS-invariant, but then $Y(\chi,\Chi)$ is independent of $\Chi$ and therefore $Z=0$. We conclude that this approach for constructing KMS-invariant actions works only if the seed expression $X$ is quadratic in $\Chi$ and $\cT$-even. The resulting action $I_d[\chi,\Chi]=\int d^4 x\, Z(\chi,\Chi)$ contains both quadratic and linear in $\Chi$ terms whose coefficients are correlated, ensuring the fluctuation-dissipation theorem. We will refer to an action $I_d$ constructed in this manner as the dissipative term. The most general SK action is a sum of $I_d$ and a non-dissipative term $I_{nd}$.

There is another type of terms satisfying the requirements 1-4 as well as KMS-invariance. Let $S[\chi]$ be any real $\cT$-invariant action. Then 
\begin{equation}\label{Indgeneral}
I_{nd}[\chi,\Chi]=\int d^4x\, \Chi(x) \frac{\delta S}{\delta\chi(x)}  
\end{equation}
clearly satisfies the conditions 1-4. It is also easy to check that it is KMS-invariant up to a total derivative. We will refer to such terms as non-dissipative, since they are real and their equations of motion are simply the Euler-Lagrange equations for $S[\chi].$

To illustrate how this procedure works in practice, let us derive a simple Schwinger-Keldysh EFT for superconductors in the London limit where the only degree of freedom is a periodic scalar $\phi$ which is proportional to the phase of the BCS condensate. We normalize it so that under an electromagnetic $U(1)$ transformation it transforms as $\varphi\mapsto\varphi+\alpha$. Let $\qphi$ be its SK partner field. The temperature is assumed constant for simplicity (this assumption will be relaxed in Section \ref{EFTI}). The diagonal $U(1)$ transformations are
\begin{equation}
\phi\mapsto\phi+\alpha, \quad \qphi\mapsto\qphi,
\end{equation}
the anti-diagonal $U(1)$ transformations are
\begin{equation}
\phi\mapsto\phi,\quad \qphi\mapsto\qphi+\alpha'.
\end{equation}

Under the conventional time-reversal symmetry, both $\phi$ and $\qphi$ are $\cT$-odd. More precisely, $\cT(\phi)=-\cTT(\phi),$ where $\cTT$ is the operator acting on a functions of $x^0$ which negates the argument $x^0$: $\cTT(f)(x^0)=f(-x^0)$. Therefore the KMS transformations are
\begin{equation}\label{KMSLondon}
\phi\mapsto -\cTT(\phi), \quad
\qphi \mapsto -\cTT(\qphi + i T_0^{-1} \partial_0\phi).
\end{equation}
In accordance with the Josephson relations, we identify $\mu=\partial_0\varphi$ as the local chemical potential. 

The non-dissipative terms are constructed from a real $U(1)$-invariant $\cT$-even action function $S[\phi]=-\int \Omega[\phi] d^4x$. Expanding in the number of spatial derivatives we have
\begin{equation}
\Omega[\phi]=\Omega_0(\mu)+\frac12\gamma_{jk}(\mu)\partial_j\phi\partial_k\phi+\ldots
\end{equation}
where $\Omega_0(\mu)$ is an arbitrary function and dots denote terms with more than two derivatives. The term linear in spatial derivatives of $\phi$ does not appear because it is  $\cT$-odd. 

The dissipative terms are constructed from the seed expression $X=iT_0 \sigma_{jk}(\mu)\partial_j\qphi\partial_k\qphi$ which gives
\begin{equation}
Z=-\sigma_{jk}(\mu)\partial_j\qphi\partial_k\mu+iT_0 \sigma_{jk}(\mu)\partial_j\qphi\partial_k\qphi
\end{equation}
The complete SK Lagrangian is
\begin{equation}
\cL_{SK}=\left(n_0(\mu)-\frac12\frac{\partial\gamma_{jk}}{\partial\mu}\partial_j\phi\partial_k\phi\right)\partial_0\qphi-\gamma_{jk}\partial_j\qphi\partial_k\phi-\sigma_{jk}\partial_j\qphi\partial_k\mu+iT_0 \sigma_{jk}\partial_j\qphi\partial_k\qphi .
\end{equation}
Here $n_0(\mu)=-\frac{\partial\Omega_0}{\partial\mu}$. The conditions 1-4 are satisfied provided the matrix $\sigma$ is positive. 

The physical meaning of $\gamma_{jk}$ and $\sigma_{jk}$ is revealed when one writes down the equations of motion for $\qphi$:
\begin{equation}\label{eom-simple}
\partial_0\left(n_0-\frac12\frac{\partial\gamma_{jk}}{\partial\mu}\partial_j\phi\partial_k\phi\right)=-\partial_j J_j,
\end{equation}
where 
\begin{equation}\label{current}
J_j = -\gamma_{jk} \partial_k\phi -\sigma_{jk}\partial_k\mu 
+ 2iT_0\sigma_{jk}\partial_k\qphi.
\end{equation}
Eq. (\ref{eom-simple}) has the form of a local conservation law and  expresses the conservation of particle number. The particle density is
\begin{equation}\label{J0super}
J_0=n_0-\frac12\frac{\partial\gamma_{jk}}{\partial\mu}\partial_j\phi\partial_k\phi,
\end{equation} 
while $J_j$ is the current. The expression for $J_j$ contains the non-dissipative London term $-\gamma_{jk}\partial_k\phi$, the diffusive term $-\sigma_{jk}\partial_k\mu$, and the noise term proportional to the noise field $\qphi$. Thus $\gamma_{jk}$ is the superfluid density tensor while $\sigma_{jk}$ is the Ohmic conductivity tensor. The noise term is imaginary, but can be made real by redefining $\qphi$ to be imaginary.\footnote{Within classical theory, it is natural to redefine $\qphi\mapsto -i\qphi$
 from the start. Then both the action and the KMS transformations become real.} 

Note that the seed we used to generate the dissipative terms in the action does not contain time derivatives of $\qphi$. While such terms are allowed, they are of higher order in the derivative expansion compared to the ones we included. For example, including a term $(\partial_0\qphi)^2$ in the seed would give a contribution to $J_0$ which is proportional to $\partial_0\mu$. 
 
The particle number density and the particle number current can be obtained from $\cL_{SK}$ as
\begin{equation}
J_0=\frac{\partial \cL_{SK}}{\partial(\partial_0\qphi)},\quad J_j=\frac{\partial \cL_{SK}}{\partial(\partial_j\qphi)}.
\end{equation}
That is, the physical charge and current are the Noether charge and current for the anti-diagonal symmetry. This is a general feature of the Schwinger-Keldysh formalism \cite{GloriosoLiu}.

The normal phase hydrodynamics is obtained by setting $\gamma_{jk}=0$. Equivalently, in the normal phase the SK action is allowed to depend on $\phi$ only through $\mu=\partial_0\phi$. In the normal phase $J_0=n_0=-\frac{\partial\Omega_0}{\partial\mu}$. This is the usual thermodynamic relation between particle number density and the ``grand potential'' $\Omega_0=F-\mu n_0$ where $F$ is Helmholtz free energy per unit volume.\footnote{$\Omega$ is sometimes called the Landau potential.} In the superconducting phase, one can rewrite Eq. (\ref{J0super}) in a similar manner:
\begin{equation}
J_0=-\frac{\partial\Omega}{\partial\mu},
\end{equation}
but for a grand potential that depends also on $\nabla\phi$:
\begin{equation}\label{Omegas}
\Omega(\mu,\nabla\phi)=\Omega_0(\mu)+\frac12\gamma_{jk}(\mu)\partial_j\phi\partial_k\phi.
\end{equation}
Stability requires $\Omega$ to be minimized at equilibrium, therefore $\gamma_{jk}$ must be positive. Unlike the positivity of $\sigma_{jk}$, this does not follow from the conditions 1-4.

To include a background electromagnetic field, we  promote  $U(1)_D$ and $U(1)_A$ to gauge symmetries, so that  $\alpha,\alpha'$ become arbitrary functions of $(x^0,\bx)$. The $U(1)_D$ gauge field $(a_0,\ba)$ is the physical electromagnetic field and transforms as usual: $a_0\mapsto a_0+\partial_0 \alpha$, $\ba\mapsto \ba+\nabla\alpha$. The $U(1)_A$ gauge field $(A_0,\bA)$ is the SK-partner of the physical electromagnetic field. Even though it is a background field, one cannot set it to zero at the outset, because this choice is not preserved by KMS symmetry. Indeed, the standard KMS transformations are
\begin{align}
& a_0\mapsto \cTT(a_0), & A_0\mapsto \cTT(A_0+iT_0^{-1}\partial_0 a_0),\\
& \ba\mapsto -\cTT(\ba), & \bA\mapsto -\cTT(\bA+iT_0^{-1}\partial_0 \ba).
\end{align}
These transformations do not commute with the physical $U(1)_D$ gauge symmetry, but this can be rectified by defining a new KMS symmetry which is the composition of the standard KMS symmetry and a $U(1)_A$ gauge transformation with a parameter $\alpha'=-i T_0^{-1} a_0$. The resulting KMS transformation is $U(1)_D$-covariant: 
\begin{align}\label{KMSa0cov}
& a_0\mapsto \cTT(a_0), & A_0\mapsto \cTT(A_0),\\
& \ba\mapsto -\cTT(\ba), & \bA\mapsto -\cTT(\bA+iT_0^{-1}\be),
\end{align}
where $\be=-\nabla a_0 + \partial_0\ba$ is the physical electric field. We see that we can consistently set $A_0=0$, but to ensure KMS invariance one must allow $\bA$ to be arbitrary.

The above redefinition also affects the KMS transformations of other fields by making them $U(1)_D$-covariant: \begin{align}\label{KMSphicov}
& \phi \mapsto -\cTT(\phi), & \qphi\mapsto -\cTT(\qphi+i T_0^{-1} D_0\phi),
\end{align}
where $D_0\phi=\partial_0\phi-a_0$. Note that $\partial_0\phi$ is now interpreted as the local electrochemical potential per unit charge, while the gauge-invariant quantity $\mu=D_0\phi$ is the local chemical potential. 

The non-dissipative part of the Schwinger-Keldysh action is given by
\begin{equation}
I_{nd}= -\int d^4x\left[D_j\qphi \frac{\partial\Omega[\phi]}{\partial(D_j\phi)}+ D_0\qphi\frac{\partial\Omega[\phi]}{\partial (D_0\phi)}\right],
\end{equation}
where $\Omega[\phi]$ is assumed to be invariant under $U(1)_D$ gauge transformations and the covariant derivatives are defined by 
\begin{equation}
    D_\mu\phi = \partial_\mu\phi-a_\mu, \qquad D_\mu\Phi = \partial_\mu\Phi - A_\mu.
\end{equation}
Then it is easy to see that $I_{nd}$ is invariant under $U(1)_D\times U(1)_A$ gauge symmetry. Under KMS symmetry $I_{nd}$ changes by a total derivative (provided one also transforms the background fields $a_\mu, A_\mu$). To leading order in spatial derivatives $\Omega[\phi]$ has the form
\begin{equation}
\Omega(\mu,D_j\phi)=\Omega_0(\mu)+\frac12\gamma_{jk}(\mu) D_j\phi D_k\phi.
\end{equation}

The dissipative part of the Schwinger-Keldysh action is obtained from a ``seed'' 
\begin{equation}
X=iT_0\sigma_{jk}(\mu) D_j\qphi D_k\qphi.
\end{equation}
This gives
\begin{equation}
Z=-\sigma_{jk}(\mu)D_j\qphi (\partial_k\mu-e_k)+i T_0\sigma_{jk}(\mu) D_j\qphi D_k\qphi.
\end{equation}
The complete SK Lagrangian is
\begin{equation}\label{eq:SKtoymodel}
\cL_{SK}=\left(n_0(\mu)-\frac12\frac{\partial\gamma_{jk}}{\partial\mu}D_j\phi D_k\phi\right)D_0\qphi-\gamma_{jk}D_j\qphi D_k\phi-\sigma_{jk} D_j\qphi (\partial_k\mu-e_k)+iT_0 \sigma_{jk} D_j\qphi D_k\qphi .
\end{equation}
At this stage one may set the $U(1)_A$ gauge fields to zero and replace $D_\mu\qphi$ with $\partial_\mu\qphi$ throughout. The particle number density and current are
\begin{equation}\label{eq:currentstoy}
J_0=n_0(\mu)-\frac12\frac{\partial\gamma_{jk}}{\partial\mu}D_j\phi D_k\phi,\quad J_j=-\gamma_{jk} D_k\phi+\sigma_{jk}(e_k-\partial_k\mu)+2iT_0\sigma_{jk}\partial_k\qphi.
\end{equation}
Note that the dissipative part of the current is proportional to the effective electric field $\tilde e_k=e_k-\partial_k\mu$, ensuring that the Einstein relation is satisfied. The correct combination of $e_k$ and $\partial_k\mu$ arises from imposing the KMS symmetry. 

If the background fields are time-independent, any action satisfying the conditions 1-4 results in equations of motion which respect the local Second Law \cite{CrossleyGloriosoLiu}. That is, there exist an entropy density $s_0$ and an entropy current $s_j$ satisfying
\begin{equation}
\partial_\mu s_\mu=\partial_0 s_0+\partial_j s_j\geq 0. 
\end{equation}
For the model we are considering, they are
\begin{equation}
    s_0 = -\frac{1}{T_0}\bigg(\Omega - (\mu+a_0)\frac{\partial\Omega}{\partial\mu}\bigg),
\end{equation} \begin{equation}
    s_j = -\frac{1}{T_0}(\mu+a_0)J_j.
\end{equation}
The positive-definite entropy production rate is
\begin{equation}
    \partial_\mu s_\mu = \frac{1}{T_0}\sigma_{jk}(\partial_j\mu-e_j)(\partial_k\mu-e_k).
\end{equation}
Note that only the electric conductivity tensor $\sigma$ contributes to entropy production.

The SK Lagrangian (\ref{eq:SKtoymodel}) is equivalent to a stochastic PDE with a Gaussian noise (the Langevin equation). To see this, we introduce a new field $\hat\Phi$ and replace the last term in (\ref{eq:SKtoymodel}) with 
$$
-2T_0\sigma_{jk}\partial_j\Phi \partial_k\hPhi+iT_0\sigma_{jk} \partial_j\hPhi \partial_k\hPhi.
$$
Integrating out $\hPhi$ gives back the original SK action, so this replacement does not affect the physics. On the other hand, $\Phi$ now enters linearly and integrating it out gives a differential equation for $\phi$:
$$
\partial_0 J_0=-\partial_j \left(J_j-2T_0\sigma_{jk} \partial_k\hPhi\right),
$$
where the second term in parentheses can be interpreted as a stochastic contribution to the current. The noise $\hPhi$ has a Gaussian distribution, with the 2-point function
\begin{equation}\label{eq:hPhicorr}
\langle\hPhi({\bf x},t)\hPhi({\bf x'},t')\rangle=\frac{\delta(t-t')}{2T_0}\left(-\sigma_{lm}\partial_l \partial_m\right)^{-1}({\bf x},{\bf x'}). 
\end{equation}
Taking into account the form of $J_j$ (\ref{eq:currentstoy}), one can also interpret the stochastic term as arising from a randomly fluctuating electric field $\hat e_k$, with the 2-point function
$$
\langle {\hat e_j}({\bf x},t){\hat e}_k({\bf x'},t')\rangle=-2T_0 \delta(t-t') \partial_j \partial_k \left(-\sigma_{lm}\partial_l\partial_m\right)^{-1}({\bf x},{\bf x'}).
$$
In the 1d case, this is a slightly disguised version of the Nyquist formula.

\section{EFT-I: Ginzburg-Landau Hydrodynamics}
\label{EFTI}

In this section we construct EFT-I: a Schwinger-Keldysh EFT for superconductors which applies near $T=T_c$. It can be viewed as a   generalization of both the normal phase SK-EFT briefly discussed in Section \ref{sec2} and the static Ginzburg-Landau theory. Thus it includes both a real scalar $\phi(x^0,\bx)$ (which describes quasi-particle excitations) and the complex Ginzburg-Landau order parameter $\psi(x^0,\bx)$. There are also their SK partners $\qphi$ and $\qpsi$. We will assume that $\psi,\qpsi$ have charge $q$ (the physical value being $q=2$). The temperature $T_0$ is still assumed to be constant (we will discuss in the next section how to remove this restriction). 
Let us list the transformations properties of the fields under all the symmetries of interest. $U(1)_D$ acts in an obvious manner:
\begin{align}
\delta_D a_\mu &=\partial_\mu \alpha, & \delta_D A_\mu &=0,\\
\delta_D\phi & =\alpha, &\delta_D\qphi &=0,\\
\delta_D\psi &=iq\alpha\psi, & \delta_D\qpsi &=iq\alpha\qpsi, \\
\delta_D\psi^* & =-iq\alpha\psi^*,   &\delta_D\qpsi^* & =-iq\alpha\qpsi^*.
\end{align}

The action of $U(1)_A$ is derived in Appendix \ref{appA}:
\begin{align}\label{u1ATDGLa}
\delta_A a_\mu  & =0, & \delta_A A_\mu &=\partial_\mu\alpha',\\
\label{u1ATDGLphi}
\delta_A \phi &=0, & \delta_A\qphi &=\alpha',\\
\label{u1ATDGLpsi}
\delta_A\psi &=0,  &\delta_A\qpsi &=iq\alpha'\psi,\\
\delta_A\psi^* & =0, &\delta_A\qpsi^* & =-iq\alpha'\psi^*.
\label{u1ATDGLpsistar}
\end{align}
The covariant derivatives are
\begin{align}
D_\mu\phi &=\partial_\mu\phi-a_\mu, & D_\mu\qphi &=\partial_\mu\qphi-A_\mu,\\
D_\mu\psi &= \partial_\mu\psi-iq a_\mu\psi, & D_\mu\qpsi &=\partial_\mu\qpsi-iqa_\mu\qpsi-i q A_\mu\psi,\\
D_\mu\psi^* &= \partial_\mu\psi^*+iq a_\mu\psi^*, & D_\mu\qpsi^* &=\partial_\mu\qpsi^*+iqa_\mu\qpsi^*+i q A_\mu\psi^*,
\end{align}
The covariant derivatives of $\phi$ and $\qphi$ are gauge-invariant, while the covariant derivatives of the fields $\psi,\psi^*,\qpsi,\qpsi^*$ transform as the respective fields.
The KMS symmetry is (see Appendix \ref{appA} for details): 
\begin{align}\label{KMSTDGLa0}
a_0 &\mapsto \cTT(a_0), & A_0 &\mapsto \cTT(A_0),\\
\label{KMSTDGLba}
\ba &\mapsto -\cTT(\ba), & \bA &\mapsto -\cTT(\bA+i T_0^{-1} \be),\\
\label{KMSTDGLphi}
\phi &\mapsto -\cTT(\phi), & 
\qphi &\mapsto -\cTT(\qphi + i T_0^{-1} D_0\phi),\\
\label{KMSTDGLpsi}
\psi&\mapsto \cTT(\psi^*), & \qpsi&\mapsto \cTT(\qpsi^*+i T_0^{-1}D_0\psi^*),\\
\psi^*&\mapsto \cTT(\psi),
& \qpsi^*&\mapsto \cTT(\qpsi+i T_0^{-1}D_0\psi).
\label{KMSTDGLpsistar}
\end{align}
The Schwinger-Keldysh action has the form $I_d+I_{nd}$. The non-dissipative term $I_{nd}$ is linear in the fluctuating fields $\qpsi,\qpsi^*,\qphi$ and arises from an action $S[\psi,\psi^*,\mu]$. More explicitly, assuming $S=-\int d^4x\, \Omega$ depends only on the first derivatives of $\psi,\psi^*$ and only on the time derivatives\footnote{Allowing $\cL$ to depend on spatial derivatives of $\phi$ would mean that there there are more superfluid degrees of freedom than those described by the Ginzburg-Landau field $\psi$. Such a situation is theoretically possible if there are several kinds of particles which form Cooper pairs or Bose-condense at different temperatures.} of $\phi$, we have:
\begin{equation}
I_{nd}= -\int d^4x \left[\left(\qpsi\frac{\partial \Omega}{\partial\psi}+D_0\qpsi\frac{\partial \Omega}{\partial(D_0\psi)}+D_j\qpsi\frac{\partial \Omega}{\partial(D_j\psi)}+c.c\right)
+D_0\qphi\frac{\partial \Omega}{\partial\mu}\right].
\end{equation}
If $\Omega$ is invariant under $\delta_D$, then it is easy to see that $I_{nd}$ is invariant under both $\delta_D$ and $\delta_A$. As for the KMS symmetry,  $I_{nd}$ is
invariant up to a total time-derivative: 
\begin{equation}
I_{nd}\mapsto I_{nd}-\frac{i}{T_0}\int d^4x \, \partial_0\Omega[\psi,\psi^*,\mu].
\end{equation}

Near the phase transition, we can expand $\Omega$ in powers of $\psi,\psi^*,\mu$ and their derivatives. 
To make this precise, we define a suitable power-counting scheme by assigning weights to fields and derivatives. The fields $\psi$, $\psi^*$, and the spatial derivative $\nabla$ have weight $1$, so that both  $|\psi|^4$ and $|\nabla\psi|^2$ have weight $4$. The term $|\psi|^2$ has weight $2$, so a consistent power-counting scheme which treats $|\psi|^2$ and $|\psi|^4$ in $\cL$ as being comparable requires us to assign weight $2$ to the coefficient $\alpha$ of $|\psi|^2$. This reflects the fact that in Ginzburg-Landau theory $\alpha\simeq a\cdot ( T_0-T_c)$, and we assume $|T_0-T_c|\ll T_0$.  Then the leading (weight $4$) terms in $\Omega$ are
\begin{equation}
\Omega=-\frac{i\Gamma_2}{2}\left(\psi^*D_0\psi-\psi D_0\psi^*\right)+\frac{1}{2}\lambda_{jk} D_j\psi^* D_k\psi+\alpha|\psi|^2+\frac{\beta}{2}|\psi|^4 - \frac12\zeta\mu^2.
\end{equation}
The coefficients $\Gamma_2,\lambda_{jk},\alpha,\beta,\zeta$ are real constants. Since the normalization of the field $\psi$ is not physical, one may let $\beta=1$ without a  loss of generality.  The corresponding contribution to the Schwinger-Keldysh action is
\begin{multline}
I_{nd}=\int d^4x \left[ \frac{i\Gamma_2}{2}\left(\qpsi^*D_0\psi-\qpsi D_0\psi^*\right)-\frac{i\Gamma_2}{2}\left(\psi D_0\qpsi^*-\psi^* D_0\qpsi\right)-\frac12\lambda_{jk}D_j\qpsi^* D_k\psi \right.\\ 
\left.
-\frac12\lambda_{jk} D_j\qpsi D_k\psi^* -\alpha(\qpsi^*\psi+\psi^*\qpsi)-\beta(\qpsi^*\psi+\psi^*\qpsi) |\psi|^2+\zeta\mu D_0\Phi \right]. \label{actionnd}
\end{multline}

The dissipative term $I_d$ is constructed from a ``seed'' $X$ which is $i$ times a positive-definite homogeneous quadratic function of $\qpsi,\qpsi^*,\qphi$ invariant under $\cT$ and $U(1)_D\times U(1)_A$. At leading order in the power counting it is easy to see that the most general such expression is a sum of two terms:
\begin{equation}\label{TDGLLOseed}
X=2iT_0\Gamma_1 |\qpsi-iq \qphi\psi|^2+iT_0\sigma_{jk} D_j\qphi D_k\qphi,
\end{equation}
where $\Gamma_1>0$ and $\sigma_{jk}$ is a positive-definite matrix. Thus there are two dissipative transport coefficients at this order. The corresponding contribution to the SK action is
\begin{multline}
I_d =-\Gamma_1\int d^4x\left[(\qpsi^*+iq\qphi\psi^*)(D_0\psi-iq\mu\psi)+(\qpsi-iq\qphi\psi)(D_0\psi^*+iq\mu\psi^*)\right. \\ \left.
-2i T_0|\qpsi-i q\qphi\psi|^2\right]
-\sigma_{jk}\int d^4x \left[ D_j\qphi (\partial_k\mu-e_k)-i T_0 D_j\qphi D_k\qphi\right]. \label{actiond}
\end{multline}
At this stage we can set $A_\mu=0$. 

To convert the theory to the standard ``Langevin'' form, we introduce new fields $\hPsi,\hPsi^*,\hPhi$ and replace the terms in $I_d$ which are quadratic in the fields $\Psi,\Psi^*,\Phi$ with
$$
\int d^4x \left[-2T_0 \Gamma_1\left((\Psi^*+iq\Phi\psi^*)\hPsi+\hPsi^*(\Psi-iq\Phi\psi)\right)+2iT_0\Gamma_1 \hPsi^*\hPsi-2T_0\sigma_{jk}\partial_j\Phi\partial_k\hPhi+i T_0\sigma_{jk}\partial_j\hPhi\partial_k\hPhi\right].
$$
Since integrating out $\hPsi,\hPsi^*,\hPhi$ gives back the original action, this does not change the physical content of the theory. Now the fields $\Psi,\Psi^*$ enter linearly and integrating them out gives Langevin equations for $\psi,\psi^*,\phi$. They read:
\begin{align}
\Gamma_1 (D_0\psi-iq\mu\psi)-i\Gamma_2 D_0\psi&=\frac12\lambda_{jk}D_j D_k\psi-\alpha\psi-\beta\psi|\psi|^2-2T_0\Gamma_1 \hPsi, \label{genLangevin1}\\
\Gamma_1 (D_0\psi^*+iq\mu\psi^*)+i\Gamma_2 D_0\psi^*&=\frac12\lambda_{jk}D_j D_k\psi^*-\alpha\psi^*-\beta\psi^*|\psi|^2-2T_0\Gamma_1 \hPsi^*, \label{genLangevin2}
\end{align}
\begin{multline}
\zeta\partial_0\mu =-q\Gamma_1\left(i\psi^*(D_0\psi-iq\mu\psi)+c.c\right)+\sigma_{jk}\partial_j(\partial_k\mu-e_k)\\
+2T_0\sigma_{jk}\partial_j\partial_k\hPhi-2iq T_0\Gamma_1 (\psi^*\hPsi-\hPsi^*\psi).
\end{multline}
Using the first two equations, the last one can be re-written as a conservation equation:
\begin{equation}\label{conservationTDGL}
\partial_0 (\zeta\mu-q\Gamma_2|\psi|^2)=-\partial_j J_j,
\end{equation}
where the particle number current is
\begin{equation}\label{currentTDGL}
J_j=q\lambda_{jk}\frac{i}{2}\left(\psi^* D_k\psi-\psi D_k\psi^*\right)+\sigma_{jk}(e_k-\partial_k\mu)-2T_0\sigma_{jk}\partial_k\hPhi.
\end{equation}
Thus the particle number density is $J_0=\zeta\mu-q\Gamma_2|\psi|^2$, and the parameter $\zeta$ can be identified with the charge compressibility of the normal component of the conducting ``fluid''.
The three terms in the current (\ref{currentTDGL}) are the superfluid contribution, the quasi-particle contribution, and the Nyquist-Johnson noise. The noise field $\hPhi$ is Gaussian and has the 2-point function (\ref{eq:hPhicorr}), the noise fields $\hPsi,\hPsi^*$ are also Gaussian, with the 2-point function
$$
\langle\hPsi({\bf x},t)\hPsi({\bf x'},t')\rangle=\frac{1}{2T_0\Gamma_1}\delta(t-t')\delta^3({\bf x}-{\bf x'}).
$$


Let us compare this with TDGL equations. For simplicity, we set the noise fields $\hPsi,\hPsi^*,\hPhi$ to zero. Then it is easy to see that Eqs. (\ref{genLangevin2}-\ref{currentTDGL}) reduce to the Schmid-Gor'kov-Eliashberg equations if we set $\Gamma_2=0$. In fact, it was noticed in \cite{Schmid} (see also \cite{LVreview}) that the parameter $\Gamma_1$ can be complexified. Here we see that simply replacing $\Gamma_1$ with a complex parameter  $\Gamma_1-i\Gamma_2$ does not give the correct equations.

In the SK formalism the existence of the entropy density and entropy current with a positive entropy production is automatic. The entropy density and current are 
\begin{align}
    s_0 &=\frac{1}{T_0}\left(-\Omega + (\mu+a_0)\frac{\partial\Omega}{\partial\mu} + (D_0\psi+iqa_0\psi)\frac{\partial\Omega}{\partial(D_0\psi)} + (D_0\psi^*-iqa_0\psi^*)\frac{\partial\Omega}{\partial(D_0\psi^*)}\right) \nonumber \\
    &= -\frac{1}{T_0}\left(q\mu\Gamma_2\abs{\psi}^2 + \frac{1}{2}\lambda_{jk}D_j\psi D_k\psi^* + \alpha \abs{\psi}^2 + \frac{\beta}{2}\abs{\psi}^4 -\frac{1}{2}\zeta \mu^2 +(\mu+a_0)J_0\right),
\end{align}
\begin{align}
    s_j &= \frac{1}{T_0}\bigg(-(\mu+a_0)\frac{\partial\cL_d}{\partial(D_j\Phi)} + (D_0\psi+iqa_0\psi)\frac{\partial\Omega}{\partial(D_j\psi)}+ (D_0\psi^*-iqa_0\psi^*)\frac{\partial\Omega}{\partial(D_j\psi^*)}\bigg) \nonumber \\
    &= \frac{1}{T_0}\left(\frac{1}{2}\lambda_{jk}((D_0\psi-iq\mu\psi) D_k\psi^* + (D_0\psi^*+iq\mu\psi^*) D_k\psi) - (\mu+a_0)J_j\right).
\end{align}
Then the entropy production rate is
\begin{equation}
    \partial_\mu s_\mu = \frac{1}{T_0}2\Gamma_1 (D_0\psi - iq\mu\psi)(D_0\psi^*+iq\mu\psi^*) + \frac{1}{T_0}\sigma_{jk}(\partial_j\mu - e_j)(\partial_k\mu-e_k). \label{entropyprodrate}
\end{equation}
A similar calculation was performed by Schmid \cite{Schmid} in the special case $\Gamma_2=0$.

We stress that the expressions for entropy density and current are valid only when the background fields $a_\mu$ are time-independent. Only in this case is the entropy production rate $\partial_\mu s_\mu$  positive-definite and hence the local Second Law follows from KMS invariance. 

In Appendix \ref{appA}, we perform the hyrodynamic expansion to the next-to-leading order. Apart from obvious modifications, such as a term $|\psi|^6$ in the potential energy and the dependence of various linear order coefficients on $\mu$ and $|\psi|^2$, we find exactly one new dissipative transport coefficient. The corresponding modification of the TDGL equations appeared in \cite{Schmid,Gulian}.

\section{EFT-II and Effects of Inhomogeneous Temperature} \label{EFTII}

In this section we show how to include non-uniform  temperature and heat conduction into the SK-EFT formalism. The first step is to promote the parameter $T_0$ to a dynamical field $T(x^0,\bx)$ which takes values in positive real numbers \cite{Jain}. Then the KMS transformations of a generic field $\chi$ and its SK-partner $\Chi$ become 
\begin{align}\label{KMSgeneralEuler}
&R_{KMS}(\chi)=\cT(\chi),\\
&R_{KMS}(\Chi)=\cT(\Chi+iT^{-1} D_0\chi).
\end{align}
A covariant derivative appears on the r.h.s. because we allowed for background gauge fields. However, after such a replacement our recipe for constructing a KMS-invariant non-dissipative action no longer works. 

A natural way around this difficulty is to define a new dynamical field $\tau(x^0,\bx)=\int^{x_0} T(u,\bx) du$ so that $T=\partial_0\tau$. Since $T$ is assumed positive, for a fixed $\bx$ the map $x^0\mapsto\tau$ is one-to-one, and one can invert the relation between $x^0$ and $\tau$. Let us denote the inverse function $\theta(\tau,\bx)$. Fields like $\chi$ and $\Chi$ can then be regarded as functions of $\tau,\bx$. Their KMS transformations take the form
\begin{align}\label{KMSgeneralLagrange}
&R_{KMS}(\chi)=\cT(\chi),\\
&R_{KMS}(\Chi)=\cT(\Chi+i D_\tau\chi),
\end{align}
which allow us to use the recipe from Section \ref{sec2} to construct a KMS-invariant non-dissipative action. Here we make use of the definition $a_\tau = T^{-1}a_0$ so that $D_\tau\chi = \partial_\tau\chi-a_\tau = T^{-1}D_0\chi$. Along with the field $\theta(\tau,\bx)$, one needs to introduce its SK partner $\qtheta(\tau,\bx)$. Their KMS transformations have the standard form:
\begin{align}\label{KMSthetaLagrange}
\theta &\mapsto -\cTT(\theta), 
&\qtheta &\mapsto -\cTT\left(\qtheta+i\partial_\tau\theta\right).
\end{align}

In what follows we will refer to $\tau$ as a ``proper time'' to distinguish it from the coordinate time $x^0$. For the purposes of this paper, using $\tau$ instead of $x^0$ as the time variable may be regarded as a mathematical trick which helps one to construct KMS-invariant actions. Once the action has been constructed, we will change the variables back. However, the ability to use either $x^0$ or $\tau$ as an independent time variable has a physical significance which is discussed further in Appendix \ref{appB}. Here we just make two remarks. First, treating the coordinate time $x^0$ as a dynamical field is similar to the ``Lagrangian'' method in ordinary  hydrodynamics where the spatial coordinates $x^j$ are regarded as functions of the ``reference'' coordinates $\sigma^j$ labeling material particles. At any fixed time, the map between $x^j$ and $\sigma^j$ is one-to-one, so one can equally well regard $\sigma^j$ as functions of $x^j$. In ordinary hydrodynamics this is referred to as the ``Eulerian'' method. In relativistic hydrodynamics, it is very natural (although not necessary) to supplement the spatial labels $\sigma^j$ with a proper time label $\tau$, in effect equipping each material particle with a clock. Here we see that a proper time variable is also very natural from the viewpoint of the Schwinger-Keldysh formalism, if one wants to include the effects of non-uniform temperature. For this reason we will refer to the usual approach where $x^0$ is an independent variable as the Eulerian picture, while treating  $\tau$ as an independent variable and $x^0=\theta(\tau,\bx)$ as a dynamical field will be called the Lagrangian picture. 

The second remark concerns time-translation symmetry which is present if the background fields are time-independent. In the Lagrangian picture a constant shift of the coordinate time $\theta(\tau,\bx)$ is still a symmetry, but now it also has an SK-partner symmetry which shifts $\qtheta$. Therefore we expect that the equation of motion for $\qtheta$ will enforce local energy conservation, assuming the background fields are time-independent. 

In accordance with Section \ref{sec2}, the general form of the non-dissipative action in the Lagrangian picture is \begin{equation}
I_{nd}=\int d\tau d^3x \left(\Chi \frac{\delta S}{\delta \chi}+\qtheta \frac{\delta S}{\delta\theta}+A_j \frac{\delta S}{\delta a_j}\right), \label{ndaction}
\end{equation}
where $S$ is a local functional of $\chi,\theta,$ and $a_\mu$ which is invariant under time-reversal, gauge symmetry, and constant shifts of $\theta$. To make contact with the usual transport theory, it is more convenient to re-write $I_{nd}$ in the Eulerian picture. Assuming that $S$ depends only on the first derivatives of $\chi$ and $\theta$, after some rather elaborate algebra we obtain:
\begin{multline}\label{IndgeneralEuler}
I_{nd}=-\int d^4x \left[\Chi\frac{\partial \Omega}{\partial \chi}+D_0\Chi \frac{\partial\Omega}{\partial (\partial_0\chi)}+\partial_0\qtheta \left(\Omega-T\frac{\partial\Omega}{\partial T}-\partial_0\chi\frac{\partial\Omega}{\partial(\partial_0\chi)}\right)  +D_j\Chi \frac{\partial\Omega}{\partial(D_j\chi)}\right.\\
\left.
-\partial_j\qtheta\left(\partial_0\chi\frac{\partial\Omega}{\partial (\partial_j\chi)}+T\frac{\partial\Omega}{\partial(\partial_j\tau)}\right)\right].
\end{multline}
Here $\Omega$ is defined by $S=-\int d^4x\, \Omega$. In the Eulerian picture, it is a function of $\chi,\Chi,\tau,\qtheta$ and their derivatives with respect to $x^0$ and $x^j$. We also remind that $T=\partial_0 \tau$.

Ignoring for now the dissipative part of the action, we note that the equation of motion for $\qtheta$ is a conservation law, as expected:
\begin{equation}
\partial_0 J^E_0=-\partial_j J^E_j,
\end{equation}
where 
\begin{equation}
J^E_0=\Omega-T\frac{\partial\Omega}{\partial T}-\partial_0\chi\frac{\partial\Omega}{\partial(\partial_0\chi)},\quad J^E_j=-\partial_0\chi\frac{\partial\Omega}{\partial (\partial_j\chi)}-T\frac{\partial\Omega}{\partial(\partial_j\tau)}.
\end{equation}
These expressions are completely general and apply to an arbitrary hydrodynamic EFT. In particular, if $\tau$ is the only dynamical field, the expression for the energy density appears to be the usual thermodynamic relation $u=\Omega-T\frac{\partial\Omega}{\partial T}$ between the (density of the) grand potential $\Omega$ and the (density of the) internal energy $u$. However, this formal similarity hides an important subtlety: in our formalism $\Omega$ is allowed to depend not only on $T=\partial_0\tau$, but also on $\partial_j\tau$. This is hard to interpret within the usual thermodynamics. The appearance of $\partial_j\tau$ in the grand potential is analogous to the appearance of $\partial_j\phi$ in the toy model of superconductivity considered in Section \ref{sec2}. It signals that one is dealing with a phase of matter where heat can flow non-dissipatively, just like electric current in a superconductor. Since such phases of matter have not been observed in nature, in the rest of the paper we will assume that $\Omega$ does not depend on the spatial derivatives of $\tau$. If one allows such a dependence, one obtains an exotic hydrodynamics. It is  discussed further in Appendix \ref{appC}.

Let us use this formalism to derive the fluctuating hydrodynamics of superconductors in the London limit while allowing for a non-uniform temperature. We will call it EFT-II. The only dynamical fields here are $\phi,\qphi,\theta,$ and $\qtheta$. Their KMS transformations are given by (\ref{KMSthetaLagrange}) and
\begin{align}\label{KMSphitheta}
\phi &\mapsto -\cTT(\phi), 
&\qphi &\mapsto -\cTT\left(\qphi+i\partial_\tau\phi\right).
\end{align}
The grand potential $\Omega$ may depend only on the derivatives of $\phi$ (both spatial and temporal) and $T=\partial_0\tau$. To quadratic order in spatial derivatives $\Omega$ must have the form
\begin{equation}
\Omega(\mu,T,D_j\phi)=\Omega_0(\mu,T)+\frac12\gamma_{jk}(\mu,T) D_j\phi D_k\phi.
\end{equation} 
The matrix $\gamma$ is required to be positive by thermodynamic stability.

The dissipative part of the Schwinger-Keldysh Lagrangian is constructed from a seed expression which is $i$ times a general positive quadratic function of  derivatives of $\qphi$ and $\qtheta$ 
(symmetries do not allow a  dependence on $\qphi$ or $\qtheta$ without derivatives). To leading order in spatial derivatives we can write the seed expression as
\begin{align}
X(\phi,\qphi,\tau,\qtheta) &= iT\sigma_{jk}(D_j\qphi-(\mu+a_0)\partial_j\qtheta)(D_k\qphi-(\mu+a_0)\partial_k\qtheta) \\
&\qquad +iT^2 \kappa_{jk}\partial_j\qtheta\partial_k\qtheta-2iT\eta_{jk}(D_j\qphi-(\mu+a_0)\partial_j\qtheta)\partial_k\qtheta.
\end{align}
Here $\sigma_{jk},\kappa_{jk},\eta_{jk}$ are real functions of $\mu,T$ such that the matrix function
\begin{equation}\label{transportmatrix}
    \begin{pmatrix}
    & \sigma & \eta & \\ & \eta^t & T\kappa &
    \end{pmatrix}
\end{equation}
is symmetric and positive-definite. 

Putting the dissipative and non-dissipative terms together, we get the complete Schwinger-Keldysh action for EFT-II which applies in the London limit:
\begin{multline}
I=\int d^4x\left[
-\left(\u_0+\frac12\left(\gamma_{jk}-(\mu+a_0)\frac{\partial\gamma_{jk}}{\partial\mu}-T\frac{\partial\gamma_{jk}}{\partial T}\right)D_j\phi D_k\phi\right)\partial_0\qtheta\right.\\
+\left(\nn_0-\frac12\frac{\partial\gamma_{jk}}{\partial\mu}D_j\phi D_k\phi\right) D_0\qphi 
 -\gamma_{jk}(D_j\qphi-(\mu+a_0)\partial_j\qtheta)D_k\phi \\ 
+\sigma_{jk}(D_j\qphi-(\mu+a_0)\partial_j\qtheta)(e_k-\partial_k\mu)+\kappa_{jk}\partial_j\qtheta\partial_k T-\eta_{jk}\partial_k\qtheta(e_j-\partial_j\mu)\\
-\frac{1}{T}\eta_{jk}(D_j\qphi-(\mu+a_0)\partial_j\qtheta)\partial_k T
+ iT\sigma_{jk}\left(D_j\qphi-(\mu+a_0)\partial_j\qtheta\right)\left(D_k\qphi-(\mu+a_0)\partial_k\qtheta\right)\\
\left. 
+iT^2\kappa_{jk}\partial_j\qtheta\partial_k\qtheta-2iT\eta_{jk}\left(D_j\qphi-(\mu+a_0)\partial_j\qtheta\right)\partial_k\qtheta\right].
\end{multline}
Here 
\begin{equation}
n_0(\mu,T)=-\frac{\partial\Omega_0}{\partial\mu},\quad u_0(\mu,T)=\Omega_0-T\frac{\partial\Omega_0}{\partial T}-(\mu+a_0)\frac{\partial\Omega_0}{\partial\mu}.
\end{equation}
The tensors $\gamma_{jk},\sigma_{jk},\eta_{jk},\kappa_{jk}$ are arbitrary real functions of $\mu$ and $T$. In addition, $\gamma_{jk}$ is symmetric and positive-definite, while the matrix coefficients $\sigma,\kappa,\eta$ are such that the matrix (\ref{transportmatrix}) is  symmetric and positive-definite. Note that the leading-order SK action still has an accidental symmetry under spatial inversion $\bx\mapsto -\bx$. In general, this symmetry will be broken by higher-derivative terms. We will see below that once time-reversal symmetry is dropped, spatial inversion is no longer automatic at leading order in the derivative expansion.

The equations of motion of EFT-II are conservation laws 
\begin{equation}
\partial_0 n=-\partial_j J^j,\quad \partial_0 u=-\partial_j J^E_j,
\end{equation}
where 
\begin{equation}
n=\nn_0(\mu,T)-\frac12\frac{\partial\gamma_{jk}}{\partial\mu}D_j\phi D_k\phi,\quad 
u=\u_0(\mu,T)+\frac12\left(\gamma_{jk}-(\mu+a_0)\frac{\partial\gamma_{jk}}{\partial\mu}-T\frac{\partial\gamma_{jk}}{\partial T}\right) D_j\phi D_k\phi, \label{densities}
\end{equation}
and the particle and energy currents are 
\begin{align}
&J_j=\frac{\partial\cL_{SK}}{\partial(\partial_j\qphi)} = -\gamma_{jk} D_k\phi +\sigma_{jk}(e_k-\partial_k\mu) -\frac{1}{T} \eta_{jk}\partial_k T
+ 2iT\sigma_{jk}(D_k\qphi-\mu\partial_k\qtheta) - 2iT\eta_{jk}\partial_k\qtheta.\\
&J^E_j=-\frac{\partial\cL_{SK}}{\partial(\partial_j\qtheta)}=(\mu+a_0) J_j -  \kappa_{jk}\partial_k T + \eta_{jk}(e_k-\partial_k\mu)-2iT^2\kappa_{jk} \partial_k\qtheta+ 2iT\eta_{jk}(D_k\qphi-\mu\partial_k\qtheta).
\end{align}
Apart from the ``noise'' terms proportional to derivatives of $\qphi$ and $\qtheta$, these are the expressions for the transport currents written down by Luttinger \cite{Luttinger}. The above derivation shows that the dependence of $n$ and $u$ on the superfluid velocity $\nabla\phi$ is determined by the coefficient $\gamma_{jk}$, and that the Second Law of Thermodynamics is equivalent to the positivity of the matrix (\ref{transportmatrix}). The case of the normal phase is obtained by setting $\gamma=0$, in which case $\sigma$ is the conductivity, $\kappa$ is the heat conductivity, and $\eta$ is the thermoelectric tensor. 

We can also derive the expressions for the entropy,  entropy current, and entropy production rate in the superconducting phase, valid for time-independent background fields,
\begin{equation}
    s_0 = -\frac{\partial\Omega}{\partial T},
\end{equation}
\begin{equation}
    s_j = \frac{1}{T}\left(J^E_j-(\mu+a_0)J_j\right),
\end{equation}
\begin{equation}
    \partial_\mu s_\mu = \frac{1}{T}\sigma_{jk}(e_j-\partial_j\mu)(e_k-\partial_k\mu) + \frac{1}{T^2}\kappa_{jk}\partial_jT\partial_kT-2\frac{1}{T^2}\eta_{jk}\partial_jT(e_k-\partial_k\mu).
\end{equation}

One can include the effects of inhomogeneous temperature into EFT-I in a similar way. The general expression (\ref{ndaction}) still applies, but now
$\chi$ is a collective name for any of the fields $\psi,\psi^*,\phi$, and $\Chi$ is a collective name for any of the fields $\qpsi,\qpsi^*,\qphi$. 
The grand potential $\Omega$ can be an arbitrary function of $\chi, T$ and their derivatives invariant under $U(1)_D$ gauge symmetry. To leading order in our power counting, it has the form 
\begin{equation}
\Omega= -\frac{i\Gamma_2}{2}\left(\psi^*D_0\psi-\psi D_0\psi^*\right)+\frac{1}{2}\lambda_{jk} D_j\psi^* D_k\psi+a(T-T_c)|\psi|^2+\frac{\beta}{2}|\psi|^4 - \frac12\zeta\mu^2-\frac12 BT^2.
\end{equation}
Here $B$ is a new constant related to specific heat at $T=T_c$ (see below).

The ``seed'' for the dissipative part of the action now contains additional terms:
\begin{align}
X&=2iT\Gamma_1 |\qpsi-iq \qphi\psi|^2+iT\sigma_{jk}(D_j\qphi-(\mu+a_0)\partial_j\qtheta)(D_k\qphi-(\mu+a_0)\partial_k\qtheta) \nonumber \\
&\qquad +iT^2 \kappa_{jk}\partial_j\qtheta\partial_k\qtheta-2iT\eta_{jk}(D_j\qphi-(\mu+a_0)\partial_j\qtheta)\partial_k\qtheta.
\end{align}
The corresponding dissipative part of the SK action is
\begin{multline}
I_d=\int d^4x \,\cL_d \\
= -\Gamma_1\int d^4x\left[(\qpsi^*+iq\qphi\psi^*)(D_0\psi-iq\mu\psi)+(\qpsi-iq\qphi\psi)(D_0\psi^*+iq\mu\psi^*)-2i T|\qpsi-i q\qphi\psi|^2\right]\\
+\int d^4x \bigg[ \sigma_{jk}(D_j\qphi-(\mu+a_0)\partial_j\qtheta)(e_k-\partial_k\mu)+\kappa_{jk}\partial_j\qtheta\partial_k T+\eta_{jk}(\partial_j\mu-e_j)\partial_k\qtheta\\
-\frac{1}{T}\eta_{jk}(D_j\qphi-(\mu+a_0)\partial_j\qtheta)\partial_k T
+ iT\sigma_{jk}\left(D_j\qphi-(\mu+a_0)\partial_j\qtheta\right)\left(D_k\qphi-(\mu+a_0)\partial_k\qtheta\right)\\
\left. 
+iT^2\kappa_{jk}\partial_j\qtheta\partial_k\qtheta-2iT\eta_{jk}\left(D_j\qphi-(\mu+a_0)\partial_j\qtheta\right)\partial_k\qtheta\right]. \label{gldissipative}
\end{multline}
The equations (\ref{genLangevin1}) and (\ref{genLangevin2}) are unaffected except that a constant $T_0$ is replaced with with a function $T$ and $\alpha=a\cdot (T-T_c)$. The particle number conservation equation is modified by the presence of a thermoelectric coefficient $\eta$ so that the particle number current is
\begin{equation}
J_j = \sigma_{jk}(e_k-\partial_k\mu)-\eta_{jk}\frac{1}{T}\partial_kT + q\lambda_{jk}\frac{i}{2}\left(\psi^*D_k\psi-\psi D_k\psi^*\right) + 2iT\sigma_{jk}(D_k\Phi-\mu\partial_k\Theta)-2iT\eta_{jk}\partial_k\Theta
\end{equation}
and the particle number density $J_0 = \zeta\mu - q\Gamma_2\abs{\psi}^2$ is unaffected. There is also a new equation obtained by varying $\qtheta$. It expresses conservation of energy and has the form
\begin{equation}
\partial_0 J_0^E=-\partial_j J^E_j,
\end{equation}
where
\begin{align}
J^E_0 &= \Omega-T\frac{\partial\Omega}{\partial T} - (\mu+a_0)\frac{\partial\Omega}{\partial\mu}-(D_0\psi+iqa_0\psi)\frac{\partial\Omega}{\partial(D_0\psi)}-(D_0\psi^*-iqa_0\psi^*)\frac{\partial\Omega}{\partial(D_0\psi^*)}\nonumber \\
&=\frac12\lambda_{jk} D_j\psi D_k\psi^*-a T_c |\psi|^2+\frac12\beta|\psi|^4+\frac12\zeta\mu^2+\frac12BT^2 + a_0J_0, \\
J^E_j & = (\mu+a_0)J_j-\kappa_{jk}\partial_k T + \eta_{jk}(e_k-\partial_k\mu) \nonumber \\
&\qquad-\frac12\lambda_{jk}\left((D_0\psi-iq\mu\psi) D_k\psi^* + (D_0\psi^*+iq\mu\psi^*)D_k\psi \right) -2i T^2\kappa_{jk}\partial_k\qtheta.
\end{align}
In particular, $\zeta$ is again identified as charge compressibility and $BT_c$ is the specific heat at $T=T_c$. The coefficients $\beta,\zeta,B$ and the matrix $\lambda_{jk}$ must be positive by thermodynamic stability.

Note that the dominant (weight $3$) terms in the energy current are the heat conductivity term and the thermoelectric term. If the thermoelectric coefficient is small (as is usually the case), when solving for $T(x^0,\bx)$ it is justifiable to neglect all other terms and get the usual heat conduction equation. The resulting $T(x^0,\bx)$ can be plugged into the equations for $\psi$ and $\mu$. 

The entropy density, current, and production rate, with time-independent background fields, are 
\begin{equation}
    s_0 = -\frac{\partial\Omega}{\partial T} =  -a\abs{\psi}^2 + BT,
\end{equation}
\begin{align}
    s_j &= \frac{1}{T}\left(J^E_j - (\mu+a_0)\frac{\partial\cL_d}{\partial(D_j\Phi)} + (D_0\psi+iqa_0\psi)\frac{\partial\Omega}{\partial(D_j\psi)}+ (D_0\psi^*-iqa_0\psi^*)\frac{\partial\Omega}{\partial(D_j\psi^*)}\right) \nonumber \\
    &= \frac{1}{T}\left(-\kappa_{jk}\partial_kT + \eta_{jk}(e_k-\partial_k\mu)\right),
\end{align}
\begin{align}
    \partial_\mu s_\mu &= \frac{1}{T}2\Gamma_1\left(D_0\psi-iq\mu\psi\right)\left(D_0\psi^*+iq\mu\psi^*\right) \nonumber
    \\ &\qquad + \frac{1}{T}\sigma_{jk}(e_j-\partial_j\mu)(e_k-\partial_k\mu) + \frac{1}{T^2}\kappa_{jk}\partial_jT\partial_kT + 2\frac{1}{T^2}\eta_{jk}(\partial_j\mu-e_j)\partial_kT.
\end{align}
The positivity of the entropy production is ensured by $\Gamma_1\geq 0$ and the usual inequalities on the tensors $\sigma,\kappa,\eta$.

\section{Broken Time-Reversal Invariance} \label{timereversal}

The local KMS condition can be formulated as a KMS symmetry only if the system is invariant under  time-reversal. To deal with the case of explicit or spontaneous breaking of  time reversal, one can use a simple trick: stack the system of interest with its image under time reversal. The variables of the time-reversed system will be distinguished with a prime. The composite has a time-reversal symmetry which exchanges the primed and unprimed variables. This allows one to apply the usual recipe for constructing hydrodynamic actions. There is no interaction between primed and unprimed variables, so afterwards one can simply discard the primed variables. 

To illustrate how this works, let us consider EFT-II which describes superconducting  hydrodynamics in the London limit. Given that the action of time-reversal is to swap primed and unprimed variables, we take the KMS transformations to be
\begin{align}
    \phi &\mapsto -\mathcal{T}(\phi'), & \phi'&\mapsto -\mathcal{T}(\phi), \label{eq:21}\\
    \tau &\mapsto -\mathcal{T}(\tau'), & \tau' &\mapsto -\mathcal{T}(\tau), \\
    \Theta &\mapsto -\mathcal{T}(\Theta' + i(T')^{-1}), & \Theta' &\mapsto -\mathcal{T}(\Theta + iT^{-1}), \\
    \Phi &\mapsto -\mathcal{T}(\Phi' + i(T')^{-1}\mu'), & \Phi' &\mapsto -\mathcal{T}(\Phi + iT^{-1}\mu), \\
    a_0 &\mapsto \mathcal{T}(a_0'), & a_0' &\mapsto \mathcal{T}(a_0), \\
    \vb{a} &\mapsto -\mathcal{T}(\vb{a}), & \vb{a}' &\mapsto -\mathcal{T}(\vb{a}), \\
    \vb{A} &\mapsto -\mathcal{T}(\vb{A}'+i(T')^{-1}\vb{e}'), & \vb{A}' &\mapsto -\mathcal{T}(\vb{A} + iT^{-1}\vb{e}). 
\end{align}
The non-dissipative action is constructed in exactly the same manner as before. The grand potential can be written as a sum
\begin{equation}
    \Omega_\cT = \Omega + \Omega'
\end{equation}
where $\Omega$ depends only on unprimed variables and $\Omega'$ depends only on primed variables. To quadratic order in spatial derivatives, the grand potential for the unprimed system takes the form
\begin{multline}
\Omega=\Omega_0(\mu,T)+\frac12\gamma_{jk}(\mu,T)(D_j\phi-l_j(\mu,T,\vb{e}))(D_k\phi-l_k(\mu,T,\vb{e}))-\sigma^A_{jk}(\mu,T) D_j\phi(\partial_k\mu-e_k) \\
- \kappa^A_{jk}(\mu,T)\partial_j\tau\partial_kT + \frac12T^{-1}(\eta_{kj}(\mu,T)-T\nu_{jk}(\mu,T))D_j\phi\partial_kT, \label{grandpot}
\end{multline}
where $l_j$ is an arbitrary vector-valued function of $\mu$, $T$, and the electric field $\vb{e}$ called the Lifshitz coupling. For any fixed $\mu$ and $T$ it can be removed by a redefinition of $\phi$ up to boundary effects \cite{KapRadz}, but if $\mu$ or $T$ are non-uniform, then the Lifshitz coupling can have bulk effects. Note that the Lifshitz coupling breaks spatial inversion. We have included a linear combination of two tensors $\eta$ and $\nu$ in the last term in order to make the parallel with transport theory more clear when we write down the energy and charge currents. Only the anti-symmetric parts of $\sigma^A$ and $\kappa^A$ contribute non-trivially to the action, hence the ``$A$'' superscript. 

The grand potential for the primed system $\Omega'$ takes the same form as $\Omega$ except that we substitute the quantities $\Omega_0, \gamma, \sigma^A, \kappa^A, \vb{l} , \eta, \nu$ with independent quantities labeled by a prime $\Omega'_0, \gamma', \sigma'^A, \kappa'^A, \vb{l'} , \eta', \nu'$. Our procedure for generating a KMS invariant action requires that the full grand potential $\Omega_\cT = \Omega + \Omega'$ is time-reversal invariant. This imposes the following constraints:
\begin{align}
    \Omega_0(\mu,T) &= \Omega'_0(\mu,T), & \gamma_{jk}(\mu,T) &= \gamma'_{jk}(\mu,T), \nonumber \\
    \sigma^A_{jk}(\mu,T) &= -\sigma'^A_{jk}(\mu,T), & \kappa^A_{jk}(\mu,T) &= -\kappa'^A_{jk}(\mu,T), \nonumber \\
    \eta_{kj}(\mu,T) - T\nu_{jk}(\mu,T) &= -\eta'_{kj}(\mu,T) + T\nu'_{jk}(\mu,T), & l_k(\mu,T) &= -l'_k(\mu,T). \label{tensorconstraints}
\end{align}

To construct the dissipative part of the action, we need to start with the ``seed'' that is $i$ times a positive expression which is quadratic in $\qphi,\qtheta,\qphi',\qtheta'$ and an arbitrary function of $\phi,\tau,\phi',\tau'$:
\begin{multline}
X(\phi,\qphi,\tau,\qtheta) = iT\sigma'^S_{jk}(\mu,T)(D_j\qphi-(\mu+a_0)\partial_j\qtheta)(D_k\qphi-(\mu+a_0)\partial_k\qtheta) \nonumber \\
\quad +iT^2 \kappa'^S_{jk}(\mu,T)\partial_j\qtheta\partial_k\qtheta-iT(\eta'_{kj}(\mu,T)+T\nu'_{jk}(\mu,T))(D_j\qphi-(\mu+a_0)\partial_j\qtheta)\partial_k\qtheta \nonumber \\
\quad +iT'\sigma^S_{jk}(\mu',T')(D_j\qphi'-(\mu'+a'_0)\partial_j\qtheta')(D_k\qphi'-(\mu'+a'_0)\partial_k\qtheta') \nonumber \\
\quad +iT'^2 \kappa^S_{jk}(\mu',T')\partial_j\qtheta'\partial_k\qtheta'-iT'(\eta_{kj}(\mu',T')+T'\nu_{jk}(\mu',T'))(D_j\qphi'-(\mu'+a'_0)\partial_j\qtheta')\partial_k\qtheta'.
\end{multline}
Here the tensors $\sigma^S, \kappa^S, \sigma'^S, \kappa'^S$ are given an ``S'' superscript to emphasize that only their symmetric parts contribute to the action, as was the case in Section \ref{EFTII} when we studied EFT-II with time-reversal invariance. Our procedure for generating a KMS invariant action requires that the seed is $\cT$-even. This condition relates primed and unprimed coefficients:
\begin{align}\label{tensorconstraints2}
    & \sigma^S_{jk}(\mu,T) = \sigma'^S_{jk}(\mu,T), \qquad \kappa^S_{jk}(\mu,T) = \kappa'^S_{jk}(\mu,T),\\ &\eta_{kj}(\mu,T)+T\nu_{jk}(\mu,T) = \eta'_{kj}(\mu,T)+T\nu'_{jk}(\mu,T). \label{tensorconstraints3}
\end{align}
Combining the constraints (\ref{tensorconstraints}) and (\ref{tensorconstraints2}, \ref{tensorconstraints3}), we find the usual Onsager relations between thermoelectric coefficients of the system and its time-reversal:
\begin{equation}
    \eta_{jk}(\mu,T) = T\nu'_{kj}(\mu,T), \qquad T\nu_{jk}(\mu,T) = \eta'_{kj}(\mu,T).
\end{equation}

The full action $I_\cT$ with non-dissipative and dissipative parts is generated by the usual procedure and can be written as a sum
\begin{equation}
    I_\cT = I + I'
\end{equation}
where $I$ depends only on unprimed variables and $I'$ depends only on primed variables. The action for the unprimed system takes the form
\begin{multline}
I=\int d^4x \left[\,
n D_0\qphi -u\partial_0\qtheta
 -\gamma_{jk}(D_j\qphi-(\mu+a_0)\partial_j\qtheta)(D_k\phi-l_k) \right.\\ 
+\sigma_{jk}(D_j\qphi-(\mu+a_0)\partial_j\qtheta)(e_k-\partial_k\mu)+\kappa_{jk}\partial_j\qtheta\partial_k T-\eta_{jk}\partial_k\qtheta(e_j-\partial_j\mu)\\
-\nu_{jk}(D_j\qphi-(\mu+a_0)\partial_j\qtheta)\partial_k T
+ iT\sigma_{jk}\left(D_j\qphi-(\mu+a_0)\partial_j\qtheta\right)\left(D_k\qphi-(\mu+a_0)\partial_k\qtheta\right)\\
\left. 
+iT^2\kappa_{jk}\partial_j\qtheta\partial_k\qtheta-iT(\eta_{kj}+T\nu_{jk} )\left(D_j\qphi-(\mu+a_0)\partial_j\qtheta\right)\partial_k\qtheta\right].
\end{multline}
The number and energy densites are defined as in Section \ref{EFTII}
\begin{equation}
    n = -\frac{\partial\Omega}{\partial\mu}, \qquad u = \Omega - (\mu+a_0)\frac{\partial\Omega}{\partial\mu} - T\frac{\partial\Omega}{\partial T} \label{Tdensities}
\end{equation}
using the grand potential defined by (\ref{grandpot}).
The electric conductivity tensor is the sum of a symmetric and anti-symmetric part $\sigma=\sigma^S+\sigma^A$, where $\sigma^S_{jk}=\sigma^S_{kj}$ and $\sigma^A_{jk}=-\sigma^A_{kj}$. The thermal conductivity similarly is the sum of a symmetric and an  anti-symmetric part $\kappa=\kappa^S+\kappa^A$. Note that the thermoelectric transport tensors $\eta$ and $\nu$ are independent whereas previously time-reversal required $\eta=T\nu^t$. SK constraints require that the 6$\times$6 dissipative transport coefficient matrix
\begin{equation}
    \begin{pmatrix}
 \sigma & \frac12(\eta+T\nu^t)  \\ \frac12(\eta^t+T\nu) & T\kappa 
    \end{pmatrix} \label{TM2}
\end{equation}
is positive definite.

If we now consider the unprimed system in isolation, we have a generalization of EFT-II with broken time-reversal invariance. In this model the inversion symmetry $\vb{x}\rightarrow -\vb{x}$ is no longer automatic due to the presence of the Lifshitz coupling $l_j$. The equations of motion are conservation laws $\partial_0n = -\partial_jJ_j$ and $\partial_0u = -\partial_jJ^E_j$, as before, with number and energy densities given by (\ref{Tdensities}). The number and energy currents are
\begin{multline}
J_j=-\gamma_{jk} (D_k\phi-l_k) -\sigma_{jk}(\partial_k\mu-e_k) -\frac{1}{T} \nu_{jk}\partial_k T
+ 2iT\sigma_{jk}(D_k\qphi-(\mu+a_0)\partial_k\qtheta) \\ +iT(\eta_{kj}+T\nu_{jk})\partial_k\qtheta,
\end{multline}
\begin{multline}
 J^E_j=(\mu+a_0) J_j -  \kappa_{jk}\partial_k T - \eta_{kj}(\partial_k\mu-e_k)-2iT^2\kappa_{jk} \partial_k\qtheta \\
 + iT(\eta_{kj}+T\nu_{jk})(D_k\qphi-(\mu+a_0)\partial_k\qtheta).   
\end{multline}
We see that $\eta$ and $\nu$ are the familiar thermoelectric coefficients from transport theory and the Hall effect and the thermal Hall effect are a consequence of $\sigma^A$ and $\kappa^A$ being non-zero. The Onsager reciprocity relations are not enforced in the absence of time reversal invariance and are replaced with the relations (\ref{tensorconstraints}) and (\ref{tensorconstraints2}) which are of no physical consequence when the unprimed system is considered in isolation. The entropy density and current take the usual form (valid for time-independent background fields)
\begin{equation}
    s^0 = -\frac{\partial \Omega}{\partial T}, \qquad s_j = \frac{1}{T}(J^E_j-(\mu+a_0) J_j).
\end{equation}
The entropy production rate
\begin{align} \label{eq:313}
    \partial_\mu s_\mu =& \; \frac{1}{T}\sigma_{jk}(\partial_j\mu-e_j)(\partial_k\mu-e_k) + \frac{1}{T^2}\kappa_{jk}\partial_j T\partial_k T  + \frac{1}{T^2}(\eta_{kj}+T\nu_{jk})(\partial_j\mu-e_j)\partial_kT
\end{align}
is altered accordingly. The positivity of the entropy production rate is guaranteed by the positivity of (\ref{TM2}) as a consequence of SK constraints. The Hall conductivities $\sigma^A$ and $\kappa^A$ drop out and do not contribute to entropy growth.

The same procedure can be applied to EFT-I to study broken time-reversal in Ginzburg-Landau hydrodynamics. The non-dissipative $I_{nd}$ and dissipative $I_d$ contributions to the action take the same form as in (\ref{IndgeneralEuler}) and (\ref{gldissipative}) with the appropriate terms included proportional to the Hall conductivities $\sigma^A$ and $\kappa^A$, distinct theromelectric coefficients $\eta$ and $\nu$, and a Lifshitz coupling. Transport coefficients are constrained relative to their primed partners as in (\ref{tensorconstraints}) and (\ref{tensorconstraints2}). KMS invariance provides an additional constraint on the dissipative coefficient
\begin{equation}
    \Gamma_1 = \Gamma_1'.
\end{equation}
There is no new behavior at leading order for broken time-reversal in EFT-I apart from the Hall effect and thermal Hall effect due to $\sigma^A$ and $\kappa^A$.

\section{Discussion} \label{discussion}

The main result of the paper is a derivation of the fluctuating hydrodynamics of a superconductor near $T=T_c$ entirely from symmetries and the requirement of local thermodynamic equilibrium. We showed that at leading order in the hydrodynamic expansion (which is simultaneously the leading order in $T-T_c$) the equations of motion are a version of TDGL equations with stochastic terms. If one omits the stochastic terms and sets  the non-dissipative coupling $\Gamma_2$ to zero, our TDGL equations reduce to the equations derived by Gork'kov and Eliashberg  \cite{GorkovEliashberg} from a microscopic theory of dirty superconductors. 

It is interesting to compare our version of TDGL equations (\ref{genLangevin1},\ref{genLangevin2},\ref{conservationTDGL},\ref{currentTDGL}) to other versions in literature. One difference is the presence of the non-dissipative coupling $\Gamma_2$. This coupling is often neglected because it violates the particle-hole symmetry (PHS)  $\psi\leftrightarrow\psi^*$, and within BCS theory such effects are suppressed by $T_c/E_F$, where $E_F$ is the Fermi energy. Nevertheless it is important to include $\Gamma_2$ if one is interested in phenomena which arise entirely from PHS violation, such as thermoelectricity. In addition, PHS is badly broken in the vicinity of Electronic Topological Transitions (for example, in some layered cuprates near optimal doping \cite{LVreview}).  Our analysis shows that at leading order in the hydrodynamic expansion $\Gamma_2$ is the only source of PHS violation (besides the usual thermoelectric coefficients of the normal component).

We emphasize that naive complexification of  $\Gamma_1$ (i.e. replacing $\Gamma_1$ with $\Gamma_1+i\Gamma_2$) does not give the correct TDGL equations. Naive complexification is often assumed in the TDGL literature (see e.g. \cite{EbisawaFukuyama,Dorsey,Hikamietal}) but it  may lead to incorrect conclusions. For example, Ref. \cite{Hikamietal} uses naive complexification to argue that $\Gamma_2$ is proportional to $\frac{\partial T_c}{\partial\mu}$. Our results show that at leading order in the hydrodynamic expansion $T_c$ is independent of $\mu$ regardless of the value of $\Gamma_2$. 

A closely related issue is the treatment of quasi-particle degrees of freedom. Both here and in \cite{GorkovEliashberg}, they are represented by a separate dynamical field $\phi$ or its covariant time-derivative $\mu=D_0\phi$ (in Ref.  \cite{GorkovEliashberg} the non-covariant time-derivative $\partial_0\phi$ is denoted  $\psi$). The field $\mu$ can be regarded as the electrochemical potential for quasi-particles.    
At leading order in the hydrodynamic expansion, the interaction betwen $\mu$ and the Ginzburg-Landau field $\psi$ is completely fixed by KMS symmetry and $U(1)$ gauge symmetry. In most discussions of TDGL $\mu$ is set to zero by an implicit appeal to approximate charge neutrality imposed by the long-range Coulomb interaction and an additional equation $\nabla\cdot J=0$ is imposed. However, as shown above, this procedure is justified only if $\Gamma_2=0$. In general, the charge neutrality condition is $\zeta\mu=\Gamma_2|\psi|^2$, where $\zeta$ is charge compressibility.  Substituting this back into the equation for $\psi$ and the expression for the current leads to new PHS-violating effects whose relative size is set not by the ratio $\Gamma_2/\Gamma_1$, but by  $\Gamma_2/\zeta\beta$ and $\Gamma_2\sigma/\zeta\lambda$. The latter ratio is of order $\ell/p_F\xi_0^2$, where $\ell$ is the mean free path, $\xi_0$ is the zero-temperature coherence length, and $p_F$ is the Fermi momentum. In a clean superconductor it can be much larger than $\Gamma_2/\Gamma_1\sim T_c/E_F$. 

Finally, we derived the version of the TDGL equations applicable to thermally isolated systems where the temperature may be inhomogeneous. As mentioned in the introduction, a microscopic derivation of such equations has been lacking and various phenomenological models have been used instead. Comparing our equations with those existing in the literature, we see that Model C of Ref.~\cite{hohenberghalperin} does not lead to the correct expression for the energy current, while the expressions used in \cite{UllahDorsey1,UllahDorsey2,Ussishkinetal,mukerjeehuse} are correct only if one neglects PHS-violating effects (i.e. sets $\Gamma_2=0$). In view of this, it would be interesting to re-examine fluctuation contributions to thermal conductivity and 
thermoelectric coefficients near $T=T_c$.

\begin{acknowledgments}
We are grateful to Hong Liu for sharing with us his unpublished notes on the hydrodynamics of superconductors and for comments on the draft. L. M. would like to thank Caltech's Summer Undergraduate Research Fellowship program for their hospitality. This work was supported in part by the U.S.\ Department of Energy, Office of Science, Office of High Energy Physics, under Award Number DE-SC0011632, as well as by the Simons Investigator Award.
\end{acknowledgments}


\appendix
\section{More on the Ginzburg-Landau hydrodynamics} \label{appA}

Here we provide some additional details on EFT-I. First, let us derive Eqs. (\ref{u1ATDGLa}-\ref{u1ATDGLpsistar}). Recall that in the SK formalism there are two copies of every field. Thus the Ginzburg-Landau $\psi$ becomes a pair of complex fields $\psi_1,\psi_2$ which have charge $q$ under two independent symmetries $U(1)_1,U(2)_2$ with parameters $\alpha_1,\alpha_2$. In the classical limit (which is formally the same as the $T_0\ra\infty$ limit) we have $\psi_1=\psi_2+\mathcal{O}(1/T_0)$, therefore it is convenient to introduce  $\psi=\frac12(\psi_1+\psi_2)$ and $\qpsi=\psi_1-\psi_2$ satisfying $\psi=\mathcal{O}(1),\qpsi=\mathcal{O}(1/T_0)$. Similarly, the fields $\phi,\qphi$ are expressed as $\phi=\frac12(\phi_1+\phi_2),\qphi=\phi_1-\phi_2,$ and $\qphi=\mathcal{O}(1/T_0)$. To enforce the correct scaling of the fields one may define $\mathcal{O}(1)$ fields $\tilde{\qpsi}=T_0\qpsi,\tilde{\qphi}=T_0\qphi$. The SK action expressed in terms of  $\psi,\phi,\tilde{\qpsi},\tilde{\qphi}$ depends on $T_0$ only through an overall factor $T_0$ which can be absorbed into the Planck constant. After such a redefinition, none of the symmetries can involve $T_0$.

The diagonal symmetry $U(1)_D$ is the subgroup defined by $\alpha_1=\alpha_2$, and $\psi$ and $\qpsi$ have charge $q$ with respect to it. The $U(1)_A$ symmetry has $\alpha_1=-\alpha_2=\alpha'$. Under this symmetry the fields transform as follows:
\begin{equation}
\delta_A \psi=iqT_0^{-1}\alpha'\tilde{\qpsi},\quad \delta_A\tilde{\qpsi}=iqT_0\alpha'\psi,\quad \delta_A\phi=0,\quad \delta_A\tilde{\qphi}=T_0\alpha'.
\end{equation}
For the limit $T_0\ra\infty$ to exist, we need to rescale $\alpha'\mapsto\alpha' T_0^{-1}$. Then in the $T_0\ra\infty$ limit we get
\begin{equation}
\delta_A \psi=0,\quad \delta_A\qpsi'=iq\alpha'\psi,\quad\delta\phi=0, \quad \delta_A\qphi'=\alpha'.
\end{equation}
This is equivalent to (\ref{u1ATDGLphi}-\ref{u1ATDGLpsistar}) with $\tilde\alpha=T_0^{-1}\alpha'$. Eqs. (\ref{u1ATDGLa}) are standard $U(1)$ gauge transformations.

Second, let us derive the KMS transformations (\ref{KMSTDGLa0}-\ref{KMSTDGLpsistar}). The conventional time-reversal symmetry acts on the background electromagnetic field via $a_0\mapsto \tau(a_0)$, $\ba\mapsto -\tau(\ba)$. Thus the covariant derivative $\partial_0-ia_0$ maps to $-(\partial_0+i\tau(a_0))$, while $\partial_j-ia_j$ mapsto $\partial_j+i\tau(a_j)$. In other words, $\cT$ changes the sign of $U(1)$ charge and thus must map $\psi,\qpsi$ to $\tau(\psi^*),\tau(\qpsi^*),$ respectively. The general formula (\ref{KMSgeneral}) together with the covariantization explained around Eqs. (\ref{KMSa0cov}-\ref{KMSphicov}) then implies (\ref{KMSTDGLa0}-\ref{KMSTDGLpsistar}). 

Finally, let us discuss the construction of the dissipative action $I_d$ for the Ginzburg-Landau hydrodynamics. In accordance with the general considerations in Section \ref{sec2}, $I_d$ is constructed from a seed expression $X(\psi,\qpsi,\phi,\qphi)$ which is $i$ times a positive-definite quadratic expression in $\qpsi,\qphi$ which is furthermore $\cT$-even and invariant under $U(1)_D\times U(1)_A$. Such a quadratic expression must have weight $\geq 2$. It is easy to see that the only expression which depends linearly on $\qpsi$ and is $U(1)_A$-invariant is $\Psi-q\Phi\psi$. Therefore the only $\cT$-even and $U(1)_D\times U(1)_A$-invariant expression of weight $2$ which involves $\qpsi$ is $i|\qpsi-q\qphi\psi|^2$. If only $\qphi$ is involved, then the only invariant expression of weight $2$ is $i\sigma_{jk}D_j\qphi D_k\qphi$ for some positive-definite matrix $\sigma_{jk}$. Thus the leading order ``seed'' must be given by (\ref{TDGLLOseed}).

It is easy to see that there are no weight-3 invariant ``seeds''. On the other hand, there are five weight-4 ``seeds'', namely 
\begin{equation}
i\mu|\qpsi-q\qphi\psi|^2,\quad i|\psi|^2|\qpsi-q\qphi\psi|^2,\quad i\mu D_j\qphi D_k\qphi,\quad i|\psi|^2 D_j\qphi D_k\qphi,\quad i\left|\qpsi^*\psi+\qpsi\psi^*\right|^2.
\end{equation}
Thus at next-to-leading order there appear five new dissipative coefficients (three scalar ones and two tensor ones). The effect of the first four is to allow $\Gamma_1$ and $\sigma_{jk}$ to depend linearly on $\mu$ and $|\psi|^2$. The effect of the fifth one is to add a term of the form $\psi\partial_0 |\psi|^2$ to the equation of motion (\ref{genLangevin1}). Such a correction to the ``vanilla'' TDGL equations has been previously obtained from microscopic considerations, see   \cite{AmbegaokarSchon,Gulian}. 

\section{The Lagrangian picture and the analogy between classical mechanics and thermodynamics} \label{appB}

It is well appreciated that many formulas in equilibrium thermodynamics and classical mechanics look similar. Here we explain how to sharpen this similarity by making thermodynamics a special case of classical mechanics. The key is to introduce a proper time variable $\tau$ such that a Josephson-like relation $T=\partial_0\tau$ is satisfied. Then, if one uses Eulerian variables, entropy and proper time become canonically conjugate variables. On the other hand, if one uses Lagrangian variables, then energy and coordinate time become canonically conjugate variables.

Consider a homogeneous system whose only symmetries are particle number conservation and energy conservation. Its equilibrium properties can be characterized by a grand potential $\Omega(\mu,T)$. The particle number $N$ and entropy $S$ are given by
\begin{equation}\label{NS}
N=-\frac{\partial\Omega}{\partial\mu},\quad S=-\frac{\partial\Omega}{\partial T}.
\end{equation}
To interpret these relations in mechanical terms, we introduce dynamical variables $\phi(x^0)$ and $\tau(x^0)$ such that the Josephson relations $\mu=\partial_0\phi$ and $T=\partial_0\tau$ hold and consider an action ${\mathcal S}=-\int dx^0\, \Omega(\partial_0\phi,\partial_0\tau).$ Since $\mathcal S$ is invariant under shifts of $\phi$ and $\tau$ by constants, the corresponding Euler-Lagrange equations are conservation equations:
\begin{equation}
\partial_0 N=0,\quad \partial_0 S=0, 
\end{equation}
where $N$ and $S$ are given by (\ref{NS}). $N$ and $S$ are now interpreted as momenta canonically conjugate to $\phi$ and $\tau$, respectively, with the non-trivial Poisson brackets
\begin{equation}
\{\phi,N\}=1,\quad \{\tau,S\}=1.
\end{equation}
The Hamiltonian corresponding to the action $\mathcal S$  is $U(N,S)=\mu N+ST+\Omega$, which is precisely the internal energy as defined in thermodynamics. Here it is viewed as a function of $N,S$. Since $\phi$ is a periodic variable, upon quantization $N$ becomes an operator with integral eigenvalues. On the other hand, since $S$ is not supposed to be integral, the variable $\tau$ is not periodically identified. The identification of the entropy as the variable conjugate to proper time is well-known in the action-based approach to ideal hydrodynamics.

Since $T>0$, the map between $x^0$ and $\tau$ is one-to-one. Therefore it is possible to switch to the Lagrangian picture by inverting the function $\tau(x^0)$. Now $\tau$ is viewed as an independent variable while $x^0$ is regarded as a field $\theta(\tau)$. In the Lagrangian picture the action becomes
\begin{equation}
{\mathcal S}=-\int d\tau\, log\, Z(\zeta,\beta),
\end{equation}
where $\beta=\partial_\tau\theta=1/T$, $\zeta=\partial_\tau\phi=\mu/T$, and $Z=e^{\beta\Omega}$ is the partition function. The momenta canonically conjugate to $\zeta$ and $\theta$ are
\begin{equation}
-\frac{\partial \log Z}{\partial \zeta}=N,\quad -\frac{\partial\log Z}{\partial\beta}=U.
\end{equation}
The equations of motion now read
\begin{equation}
\partial_\tau N=0,\quad \partial_\tau U=0.
\end{equation}
The non-trivial Poisson brackets are
$$
\{\phi,N\}=1,\quad \{\theta,U\}=1.
$$
Note that in the Lagrangian picture the coordinate time $x^0$ becomes a bona fide dynamical variable canonically conjugate to energy. Upon quantization the energy-time uncertainty relation follows in the standard manner. On the other hand, in the Eulerian picture one has an entropy-proper-time uncertainty relation.

\section{Superthermal Hydrodynamics}\label{appC}

There seems to be no physical principle which would prohibit the SK action in the Eulerian picture from depending on the spatial derivatives of the proper time $\tau$. We will say that such an action describes a {\it superthermal phase}. A similar phase has been previously considered by M. Liu \cite{MarioLiu}. While we are unaware of any physical realizations of such a phase, it is interesting to consider its properties.

Note first of all that in the presence of particle number  conservation there is more than one version of the superthermal phase, since the particle number symmetry may or may not be broken. To simplify the discussion, let us consider a system where the only conserved quantity is energy. Since we do not impose Galilean invariance, this setup implicitly assumes that there is also a ``substrate'' whose only role is to break invariance under Galilean boosts. In the Lagrangian picture, the SK EFT contains only two fields: the coordinate time $\theta(\tau,\bx)$ and its SK partner $\qtheta(\tau,\bx)$. The local temperature is identified as $T^{-1}=\partial_\tau \theta$. The KMS transformations are given by (\ref{KMSthetaLagrange}). The non-dissipative part of the action $I_{nd}$ is constructed from a thermodynamic potential $\Omega$ which is allowed to depend both from $T$ and $\partial_j\theta$. According to the general recipe (\ref{IndgeneralEuler}), when $I_{nd}$ is written in the conventional Eulerian picture, it takes the form
\begin{equation}
I_{nd}=-\int d^4x\left[ \partial_0\qtheta \left(\Omega-T\frac{\partial\Omega}{\partial T}\right)-T \partial_j\qtheta \frac{\partial\Omega}{\partial(\partial_j\tau)}\right].
\end{equation}
To quadratic order in spatial derivatives we must have
\begin{equation}
\Omega(T,\partial_j\tau)=\Omega_0(T)+\frac12 \zeta_{jk}(T)\partial_j\tau\partial_k\tau.
\end{equation}
The matrix $\zeta_{jk}(T)$ must be positive by thermodynamic stability. 

The dissipative part of the action is constructed from a ``seed''  which is $i$ times an expression which is quadratic in $\qtheta$, invariant under shifts of $\qtheta$ and $\tau$ by constants, and is positive. To leading order in spatial derivatives it must have the form
\begin{equation}
X(\tau,\qtheta)=iT^2\kappa_{jk}(T)\partial_j\qtheta\partial_k\qtheta,
\end{equation}
where the matrix function $\kappa_{jk}(T)$ must be positive. 
The corresponding dissipative part of the action is
\begin{equation}
I_d=\int d^4x \left(\kappa_{jk}\partial_j\qtheta\partial_k T+ iT^2\kappa_{jk}\partial_j\qtheta\partial_k\qtheta\right).
\end{equation}
The total action of the superthermal hydrodynamics is $I_{nd}+I_d$. The corresponding 
equation of motion is a local  conservation law for energy:
\begin{equation}
\partial_0 J^E_0=-\partial_j J^E_j,
\end{equation}
where
\begin{align}
J^E_0 &=\Omega-T\frac{\partial\Omega}{\partial T}=u_0(T)+\frac12\left(\zeta_{jk}-T\frac{\partial\zeta_{jk}}{\partial T}\right)\partial_j\tau\partial_k\tau,\\
J^E_j &=-T\frac{\partial\Omega}{\partial(\partial_j\tau)}-\kappa_{jk}\partial_kT-2iT^2\kappa_{jk}\partial_k\qtheta=-T\zeta_{jk}\partial_k\tau
-\kappa_{jk}\partial_kT-2iT^2\kappa_{jk}\partial_k\qtheta .
\end{align}
Here $u_0(T)=\Omega_0-T\frac{\partial\Omega_0}{\partial T}.$ 

If $\zeta_{jk}=0$, both $\Omega$ and $J_0^E=u_0(T)$ are functions of $T$ only. $\Omega$ is then interpreted as the density of the  Helmholtz free energy. This interpretation remains true even if $\zeta_{jk}$ is nonzero, but then the local state of the system is determined not only by the local temperature $T$, but also by the value of the vector $\nabla\tau$. This is analogous to the situation in a superconductor where the local state depends not only on $\mu,T$, but also on the superfluid velocity $\nabla\phi$.

The expression for the energy current is reminiscent of the London equation, as it contains, along with the usual diffusive contribution $-\kappa_{jk}\partial_k T$, a non-dissipative term $-T\zeta_{jk}\partial_k\tau.$ One can check that the entropy production is proportional to $\kappa$ but does not depend on $\zeta$.

An equivalent way to characterize the superthermal phase is to say that the heat conductivity diverges at zero frequency. Indeed, if we consider a temperature gradient which depends on time as $e^{-i\omega x^0}$, the energy  current is
\begin{equation}
J^E_j(\omega)=-\left(\frac{i}{\omega}\zeta_{jk}+\kappa_{jk}\right)\partial_k T(\omega).
\end{equation}

Finally, by analogy with superconductors, we expect that in a superthermal phase there is a gapless Goldstone mode. Intuitively, it arises from spontanenous breaking of proper time translation symmetry. To verify this, we write down the linearized  equation of motion for $\tau$ by letting $\tau(x^0,\bx) = {\bar T} x^0 + h(x^0,\bx)$ for $h(x^0,\bx)\ll 1$ and a constant $\overline T$ (the average temperature). To leading order in $h(x^0,\bx)$ and dropping the terms containing the fluctuating field $\Theta$ we get 
\begin{equation}
    -\frac{\partial^2 \Omega_0({\overline T})}{\partial {\overline T}^2}\partial_0^2h = {\overline\zeta}_{jk}\partial_j\partial_k h +\frac{{\overline\kappa}_{jk}}{{\overline T}^2}\partial_j\partial_k\partial_0 h,
\end{equation}
where ${\overline\zeta}_{jk}=\zeta_{jk}(\overline T),$ etc. 
Plugging in $h(x^0,\bx) = \text{exp}(-i\omega x^0+i k_j x_j)$ and solving for $\omega$ gives
\begin{equation}
    \omega = \pm \left(\frac{\partial^2 \Omega_0}{\partial {\overline T}^2}\right)^{-1}\sqrt{-\frac{\partial^2\Omega_0}{\partial {\overline T}^2}{\overline\zeta}_{jl }k_j k_l - \bigg(\frac{{\overline\kappa}_{jl}}{2{\overline T}^2}k_j k_l\bigg)^2} + \frac{i}{2{\overline  T}^2}\bigg(\frac{\partial^2\Omega_0}{\partial {\overline T}^2}\bigg)^{-1}{\overline\kappa}_{jl}k_j k_l.
\end{equation}
For sufficiently small $k$ this describes a propagating linearly-dispersing mode with a direction-dependent velocity tensor
\begin{equation} \label{eq:55}
    c^2(\bar T)_{jk} = \left(-\frac{\partial^2 \Omega_0}{\partial  T^2}\right)^{-1}{\overline \zeta}_{jk}.
\end{equation}
Note that the positivity of the  isochoric specific heat capacity, $C_V = -T\partial^2 \Omega_0/\partial T^2 \geq 0$ and the positivity of $\zeta_{jk}$  ensure that $c^2>0$. This mode describes waves of entropy and is similar to the second sound in ordinary superfluids.

\bibliography{Bibliography}

\end{document}